\documentclass{article}

\usepackage{arxiv}

\usepackage[utf8]{inputenc} 
\usepackage[T1]{fontenc}    
\usepackage{hyperref}       
\usepackage{booktabs}       
\usepackage{nicefrac}       
\usepackage{microtype}      
\usepackage{lipsum}

\usepackage{cite}
\usepackage{amsmath,amssymb,amsfonts}
\usepackage{algorithmic}
\usepackage{graphicx}
\usepackage{textcomp}
\usepackage{xcolor}
\usepackage{longtable}
\usepackage{booktabs,siunitx}
\usepackage{subcaption}
\usepackage{url}
\usepackage{tabularx}
\usepackage{array}
\usepackage{multirow}

\newcommand{\fourobjects}[4]{%
  \leavevmode\vbox{\hbox{#1}\hbox{#2}}\vbox{\hbox{#3}\hbox{#4}}%
}

\newcommand{\ra}[1]{\renewcommand{\arraystretch}{#1}}

\title{Assessing Post Deletion in Sina Weibo: Multi-modal Classification of Hot Topics}

\author{
  Meisam Navaki Arefi \\
  Department of Computer Science\\
  University of New Mexico\\
  Albuquerque, NM \\
  \texttt{mnavaki@unm.edu} \\
   \And
 Rajkumar Pandi \\
  Department of Computer Science\\
  University of New Mexico\\
  Albuquerque, NM \\
  \texttt{rpandi@unm.edu} \\
   \And
  Michael Carl Tschantz \\
  International Computer Science Institute (ICSI)\\
  Berkeley, CA \\
  \texttt{mct@icsi.berkeley.edu} \\
   \And
 Jedidiah R. Crandall \\
  Department of Computer Science\\
  University of New Mexico\\
  Albuquerque, NM \\
  \texttt{crandall@cs.unm.edu} \\
     \And
  King-wa Fu \\
  Journalism and Media Studies Centre\\
  University of Hong Kong\\
  Hong Kong \\
  \texttt{kwfu@hku.hk} \\
       \And
  Dahlia Qiu Shi \\
  University of Hong Kong\\
  Hong Kong \\
  \texttt{qiushi19.jlu@gmail.com} \\
       \And
  Miao Sha \\
  University of Hong Kong\\
  Hong Kong \\
  \texttt{shamiao@connect.hku.hk} \\
}

\begin{document}
\maketitle

\begin{abstract}
Widespread Chinese social media applications such as Weibo are widely known for monitoring and deleting posts to conform to Chinese government requirements. In this paper, we focus on analyzing a dataset of censored and uncensored posts in Weibo. Despite previous work that only considers text content of posts, we take a multi-modal approach that takes into account both text and image content. We categorize this dataset into 14 categories that have the potential to be censored on Weibo, and seek to quantify censorship by topic. Specifically, we investigate how different factors interact to affect censorship. We also investigate how consistently and how quickly different topics are censored.
To this end, we have assembled an image dataset with 18,966 images, as well as a text dataset with 994 posts from 14 categories. We then utilized deep learning, CNN localization, and NLP techniques to analyze the target dataset and extract categories, for further analysis to better understand censorship mechanisms in Weibo. \\
We found that \textit{sentiment} is the only indicator of censorship that
is consistent across the variety of topics we identified. Our finding matches with recently leaked logs from Sina Weibo. We also discovered that most categories like those related to anti-government actions (\emph{e.g.} protest) or categories related to politicians (\emph{e.g.} Xi Jinping) are often censored, whereas some categories such as crisis-related categories (\emph{e.g.} rainstorm) are less frequently censored. We also found that censored posts across all categories are deleted in three hours on average.
\end{abstract}


\section{Introduction}
Human monitoring of social media posts and the subsequent deletion of posts that are considered sensitive is an important aspect of Internet censorship for academic study. Seeing a post get removed by the censors gives valuable information to researchers, including the content that was censored and the amount of time it was visible before being deleted.  This information can provide insights into the censors' policies and priorities. A better understanding of censors' motivations can lead to more effective ways of addressing Internet censorship, be they technical, political, legal, economic, or otherwise.

Censorship of Chinese social media is a complex process that involves
many factors. There are multiple stakeholders and many different interests: economic, political, legal, personal, \emph{etc.}, which means that there is not a single strategy dictated by a single government authority~\cite{miller2017limits}. Moreover, sometimes Chinese social media do not follow the directives of government, out of concern that they are more strictly censoring than their competitors~\cite{miller2017limits}.

Past literature in censorship of Chinese social media has attempted to make
general statements about what kinds of features lead to a given post
being likely to be censored. Researchers have posited the topic of a
post (\emph{e.g.}, what keywords it contains)~\cite{bamman2012censorship,zhu2013velocity}, how viral or popular the
post is (\emph{e.g.}, how much it is reposted and commented on)~\cite{zhu2013velocity}, the
collective action potential (how likely it is to lead to, \emph{e.g.}, protests)~\cite{king2013censorship}, and the individual posting the content~\cite{blake2019}, as major features that determine how high of a priority deleting the post is for the censors.
However, no study to date with respect to censorship in China has considered the multimodal nature of social
media, and past studies have relied on relatively narrow datatsets
(\emph{e.g.}, spanning months rather than years or only following a
small set of users).



In this paper, we focus on Sina Weibo and use Weiboscope dataset~\cite{fu2013assessing}, which tracks 120,000 users over 4 years (2015--2018) in Sina Weibo and includes 128,044 posts, of which 64,022 were censored. The WeiboScope dataset has only two categories, censored and uncensored, and does not include the reason for censorship. In particular, this dataset is not labeled by topics and it is very time-consuming to manually categorize them. We identify fourteen topics that both (1) saw a significant amount of censorship in the WeiboScope
dataset; and, (2) could be identified through both images and text. To analyze the dataset we take a \textit{multi-modal} approach that takes into account both \emph{text} and \emph{images} that appear in posts. We then test the effect of various factors that may affect censorship that were identified by past literature on the lifetime of posts.

Sina Weibo is one of the most popular social media platforms in China (``Weibo'' means ``microblog'' in Chinese). After the Urumqi riots~\cite{roits}, Chinese authorities shut down all social media platforms including Twitter, Facebook, and local social media platforms. Sina Weibo provides microblogging services similar to Twitter but was designed to block posts with content that does not comply with the Chinese government's requirements.  
Weibo users can re-post and follow other users, mention other people with $@$UserName, and add hashtags using \#HashName\#.  More importantly for this study, Weibo also allows embedded photos.
As of July 2018, Weibo has over 441~million active users, which surpasses Twitter's 339~million active users~\cite{weibotwitter2018statistic}.

To analyze the WeiboScope dataset, we take a semi-automated multi-modal approach and utilize deep learning, CNN localization, and NLP techniques. To train our image and text classifiers, we first assembled our own image and text datasets
from 14 interesting categories that are potential topics for censorship on Weibo and any other social media platforms in China. We refer to the image dataset as \textit{CCTI14 (Chinese Censored Topics Images)}, and to the text dataset as \textit{CCTT14 (Chinese Censored Topics Text)}. 
After training classifiers with CCTI14 and CCTT14, we categorize the WeiboScope dataset into our 14 categories.

These categories are selected based on previous research, domain knowledge, and known censorship events in China. CCTI14 has 18,966 labeled images and CCTT14 has 994 labeled texts from 14 categories as well as an ``Other'' category. These categories are as follows (in alphabetical order): 1) \textit{Bo Xilai}, 2) \textit{Deng Xiaoping}, 3) \textit{Fire}, 4) \textit{Injury/Dead}, 5) \textit{Liu Xiaobo}, 6) \textit{Mao Zedong}, 7) \textit{People's congress}, 8) \textit{Policeman/Military forces}, 9) \textit{Protest}, 10) \textit{Prurient/Nudity}, 11) \textit{Rainstorm}, 12) \textit{Winnie the Pooh}, 13) \textit{Xi Jinping}, 14) \textit{Zhou Kehua}.

We trained an image classifier over the CCTI14 dataset using the VGG network~\cite{simonyan2014very} and it achieved a 97\% F1-score. We also trained a text classifier over the CCTT14 dataset that achieved a 95\% F1-score. We used our classifiers to classify both censored and uncensored posts from the target dataset under study into the above-mentioned 14 categories.  Because of a flag in the Weibo API, we can distinguish between deletions by a post's author and by the Weibo system itself, providing ground truth for which posts have been censored.  

We found that \textit{sentiment} is the only indicator of censorship that
is consistent across the variety of topics we identified. We also found that most of the categories (\emph{e.g.}, protest) are often censored, whereas some categories (\emph{e.g.}, rainstorm) are less frequently censored. This suggests that different topics can be censored with different levels of consistency.
We also found that the median lifetime of the posts that were censored in a category is less than three hours on average, which confirms that censors can quickly delete sensitive posts.

To the best of our knowledge, our work is the first to look at both text and image content of posts being censored and not just at the text content. We hope that our datasets, CCTI14 and CCTT14, which are the first datasets labeled by topics assembled for studying China's censorship, can help other researchers to uncover image and text censorship mechanisms in other social media platforms in China, and that our techniques can be applied in other contexts.



In summary, this paper presents the following contributions:
\begin{itemize}
\item We introduce CCTI14 and CCTT14, the first image and text datasets labeled by topics assembled specifically for studying image and text censorship in Chinese social media.
\item We train a CNN model over CCTI14 that achieves 97\% F1-score, and a text classifier over CCTT14 that achieves 95\% F1-score, to automatically classify the target dataset under study of this paper, based on both image and text content.
\item We use a CNN localization technique to double check that our categories and our trained image model produce an intuitive model.


\item
For each category, we analyze how quickly and how often it is censored. We also perform survival analysis per category to investigate how different factors interact to affect the lifetime of a post.

\item 
We make CCTI14, CCTT14, our code, all embedded images with and without CNN localization, and our trained models publicly available to help important efforts such as those to understand image and text censorship or to identify topics that are likely to be censored.

\end{itemize}

This paper is organized as follows. Section~2 describes the dataset under study of this paper. Section~3 explains our methods. Section~4 presents our analysis and results, and Section~5 presents related work. Finally, Section~6 concludes the paper.



\section{WeiboScope Dataset}

WeiboScope tracks about 120,000 users from three samples: 
\begin{enumerate}
\item User accounts with high going-viral potential, measured by the number of followers.
\item
A group of accounts whose posts are known to have a high likelihood to be censored, such as individual media, NGOs, human right lawyers, grassroots leaders, or key opinion leaders, etc.
\item A random sample of accounts generated by randomly selecting users' unique identity codes.
\end{enumerate}

By following the tracked users as ``friends'', the user's recently modified timeline is compared to the previous version, every 20 minutes, to discover if any posts had been deleted.
When a post is missing, Weibo returns two possible messages: ``weibo does not exist''
or ``permission denied''. The latter is returned when the censors make the post inaccessible to others, and the former message is returned when the user voluntarily deletes the post or the censors remove it entirely. Since there is no feasible way to determine who deleted a post, we only considers posts deleted by a ``permission denied'' message to be censored.


From January 2015 through April 2018, WeiboScope collected 64,022 censored 
and more than 40 million uncensored posts by tracking the above-mentioned users.
In this paper, to be able to compare censored and uncensored posts, we randomly selected 64,022 uncensored posts from the 40 million uncensored posts. We know that these posts are uncensored since they were not deleted by the censor or the user.
 Thus the reduced WeiboScope dataset that we study in this paper has 64,022 censored posts and 64,022 uncensored posts from 2015 through 2018. 


\section{Methods}

During the analysis of the target dataset we encountered a number of challenges that we present here. We also describe CCTI14 and CCTT14 datasets and our image and text classifiers to address these challenges.

\subsection{Challenges}
Here, we describe the challenges that we encountered over the course of analyzing the target dataset. 

\textbf{The possibility of interactions between multiple factors:}
To decide whether to censor a post, the censors may use any of the factors recorded in our datasets: images, text, number of reposts, number of comments, or the user account making the post.
Furthermore, censors may also use factors not recorded in our datasets, such as number of views or information about the political situation at the time.
The last possibility highlights that censorship may change over time.
Furthermore, censorship might even depend upon ideally irreverent factors, such as the motivation of a human monitor on a particular day.  

Further complicating matters, the censor may consider multiple factors at a time.
For example, an image of Xi Jinping and text sharing a proverb about the consequences of poor leadership may both be acceptable in isolation but not together.
Also, a borderline image from a sensitive user might be censored very quickly, whereas the same image from another user might not be censored at all.
Such interactions between factors pose difficulties for finding patterns that increase the likelihood of censorship.

Indeed, the target dataset includes apparently sensitive images (\emph{e.g.}, of protests) that are not censored and apparently nonsensitive images that are (\emph{e.g.}, images of a sunset).
There are even some images that appear in both censored and uncensored posts (\emph{e.g.}, images of a rainstorm). 
Such results may be produced by any of the other possible factors playing a role in the decision to censor.

\textbf{Lack of experimental data:}
Additionally, having access to observational data but not experimental data means that any found patterns may be correlated with censorship but not actually causing it.
This issue limits our abilities to draw conclusions about the causes of censorship.
While can find patterns predictive of censorship, between this limitation and the multiple possible factors discussed above, we cannot draw firm conclusions about why a post is censored.

\textbf{Lack of canonical topic categories:}
Nevertheless, given the nature of censorship, we are confident that the topics of a post's text and images, if any are present, has a causal effect on whether a post is censored.
Thus, analyzing these topics are of crucial importance to our study.
However, we lack knowledge of how censors view posts and classify them into topical categories.
For example, they may view all images of protests as simply belonging to a broad protest category, thereby increasing all of their probabilities of censorship.
Alternatively, they may have more refined categories in mind, drawing a distinction between pro- and anti-government protests, or even between particular issues being protested.

Lacking a canonical categorization of topics, we must adopt one of our own, which may lack the distinctions found in the censors' actual categorization.
This lack of refinement may appear to be noise in our data, that is, seemingly random variations in how the censor responds to the same topic.
However, in actuality, the censor may be consistently applying a rule more refined than what our analysis can identify.

\textbf{Clustering methods do not work here:} 
Lacking pre-defined categories, it may be tempting to automatically categorize the images in the target dataset with clustering algorithms. However, since the target dataset has very diverse images, clustering algorithms do not work well.
We tried several clustering algorithms (\emph{e.g.}, hierarchical and K-means), but none of them was able to cluster the images in a way that we could learn something from the categories. The clustering algorithms would either come up with: i) too many categories (where many of them have only a few images), which render the clustering useless, or ii) with a reasonable number of categories each of which contains many diverse images from which, again, nothing could be learned.


\textbf{There is no image or text dataset available for studying image and text censorship:} Furthermore, in order to be able to use ML classification methods to categorize images and texts, annotated image and text datasets are needed that is particularly designed for studying censorship in China, but there is no such datasets publicly available. 

\smallskip
To overcome these challenges, we take the very first step in collecting image and text datasets particularly for studying image and text censorship in Chinese social media. We refer to these datasets as \textit{CCTI14 (Chinese Censorsed Topics Images)} and \textit{CCTT14 (Chinese Censorsed Topics Text)}. Then we train classifiers over CCTI14 and CCTT14 to help us in categorizing image and text content of posts in the WeiboScope dataset. 


\subsection{Image Classifier}
In this section, we first describe how we assembled the CCTI14 dataset. Then we present the performance evaluation of our CNN model over CCTI14.

\subsubsection{CCTI14 Dataset}
To find a list of potentially censored categories in Weibo, we relied on previous research and censorship events in different domains of censorship in China~\cite{zhu2013velocity,king2013censorship,bamman2012censorship}. We ended up with 14 categories spanning diverse domains including collective action (e.g., Protest), Chinese politicians (e.g., Xi Jinping, Deng Xiaoping, and Mao Zedong), crisis and disaster (e.g., rainstorm and fire), political activists (e.g., Liu Xiaobo), and mockery (e.g. Winnie the Pooh). We did not include categories that we were not able to find at least 100 unique images (e.g., Xi Jinping bun) or were too vague to have them as a separate category (e.g., China anti-corruption). Our categories are not comprehensive, since there is no such comprehensive list of topics that China censors. However, we have tried to pick general categories so that they can be applied for analyzing any other Chinese platforms that practice censorship. 

\noindent\textbf{Training Dataset:}
To assemble a training dataset, we utilized Google Image Search to find images of $200 \times 200$ pixels or bigger per category. As has been done by other studies~\cite{bainbridge2013intrinsic,bainbridge2012establishing}, we scraped Google Images and automatically collected images per category. In addition to the 14 categories, we carefully crafted an ``Other'' class including random images and images that we found could be confused with other categories (e.g., street banner confused with protest and ocean confused with a rainstorm). 

As is common practice~\cite{xiao2010sun, bainbridge2013intrinsic, bainbridge2012establishing}, we then manually removed problematic images including those that were too blurry or would fall into more than one category (e.g., an image of both Deng Xiaoping and Mao Zedong). We also manually removed all duplicate images in a category or among several categories. To do so, two trained human annotators verified that images are in the right category, with each annotator spending 5 hours on average on this. In case of a disagreement between annotators about an image, an expert made a decision on the image.



We also used the label preserving image augmentation techniques to add more images to our dataset. Image augmentation is the procedure of taking images and manipulating them in different ways to create many variations of the same image. In this way, not only can we train our classifier on a larger dataset, but also we can make our classifier more robust to image coloring and noise. It has been proven that data augmentation could be very effective in improving the performance of CNNs~\cite{wong2016understanding,xu2016improved}. 

\begin{figure*}
 \centering
  \subcaptionbox{Original Image\label{fig3:a}}{\includegraphics[width=0.13\textwidth]{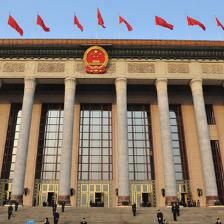}}\hspace{0.15em}%
  \hfill
  \subcaptionbox{Contrast normalization\label{fig3:b}}{\includegraphics[width=0.13\textwidth]{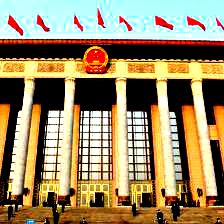}}\hspace{0.15em}%
  \hfill
  \subcaptionbox{Affine transformation\label{fig3:c}}{\includegraphics[width=0.13\textwidth]{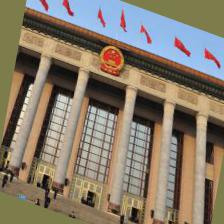}}\hspace{0.15em}%
  \hfill
  \subcaptionbox{Perspective transformation\label{fig3:c}}{\includegraphics[width=0.13\textwidth]{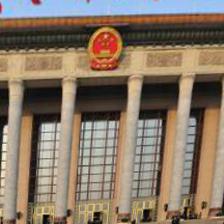}}\hspace{0.15em}%
  \hfill
  \subcaptionbox{Sharpen\label{fig3:c}}{\includegraphics[width=0.13\textwidth]{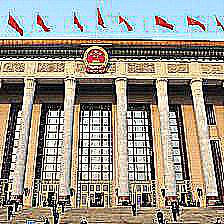}}\hspace{0.15em}%
 \hfill
    \subcaptionbox{Gaussian blur\label{fig3:c}}{\includegraphics[width=0.13\textwidth]{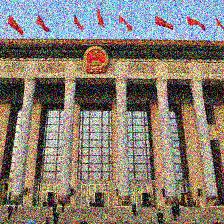}}\hspace{0.15em}%
 \hfill
  \subcaptionbox{Padding\label{fig3:d}}{\includegraphics[width=0.13\textwidth]{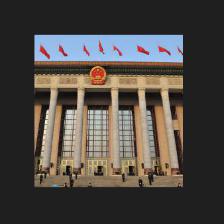}}
  \caption{An example of image augmentation.}
  \label{fig:aug}
\end{figure*}

We picked six label-preserving data augmentation techniques: i) contrast normalization, ii) affine transformation, iii) perspective transformation, iv) sharpen, v) Gaussian blur, vi) padding. We then applied them to each image in our dataset and added the result images to our dataset. Figure~\ref{fig:aug} shows an example of image augmentation in CCTI14. 


\noindent\textbf{Testing Dataset:}
The classifier should be tested against \textit{real-world} censored images from Weibo so that it can be trusted in categorizing the WeiboScope dataset which consists of real censored images.
To this end, we assembled a test image dataset from \textit{real-world} censored images.
We used two human annotators to manually label a small subset of images from WeiboScope dataset into the 15 categories. Here are the steps that we followed for assembling the testing dataset:
\begin{enumerate}
    \item We trained two human raters by providing them the definition for each category as well as image samples per category.
    \item We randomly selected 1000 censored images from WeiboScope dataset.
    \item We asked the raters to categorize these images into the 15 categories.
    \item If each category has at least 30 images, go to \#5. Otherwise go to \#2. 
    \item In case of a disagreement between raters about an image we asked an expert to categorize the image.
\end{enumerate}

At the end of this process, we measured the inter-rater reliability using Cohen's Kappa coefficient~\cite{cohen1960coefficient}. The inter-rater reliability was 91\%, which is satisfactory. Each rater spent 6 hours on average to annotate the dataset. 

The final test dataset has 1014 images 
(which is equal to about 5\% of the size of the train dataset), and each category has 30-70 images. Note that since the ``Other'' category had many more images than other categories, we only kept 70 (randomly selected) images from that category to balance the dataset.

CCTI14's training dataset has 5,038 images before augmentation, and 18,966 images after augmentation from 14 categories and one ``Other'' class in which each category has 700--1400 images. Also CCTI14's testing dataset has 1014 images from \textit{real-world} censored images from the 15 categories.

\subsubsection{CNN Model}
In this section, we present our CNN model and evaluate its performance using several metrics. We also explain how we use CNN localization for error analysis.

\begin{figure}[!ht]
\centering
      \includegraphics[width=0.55\textwidth]{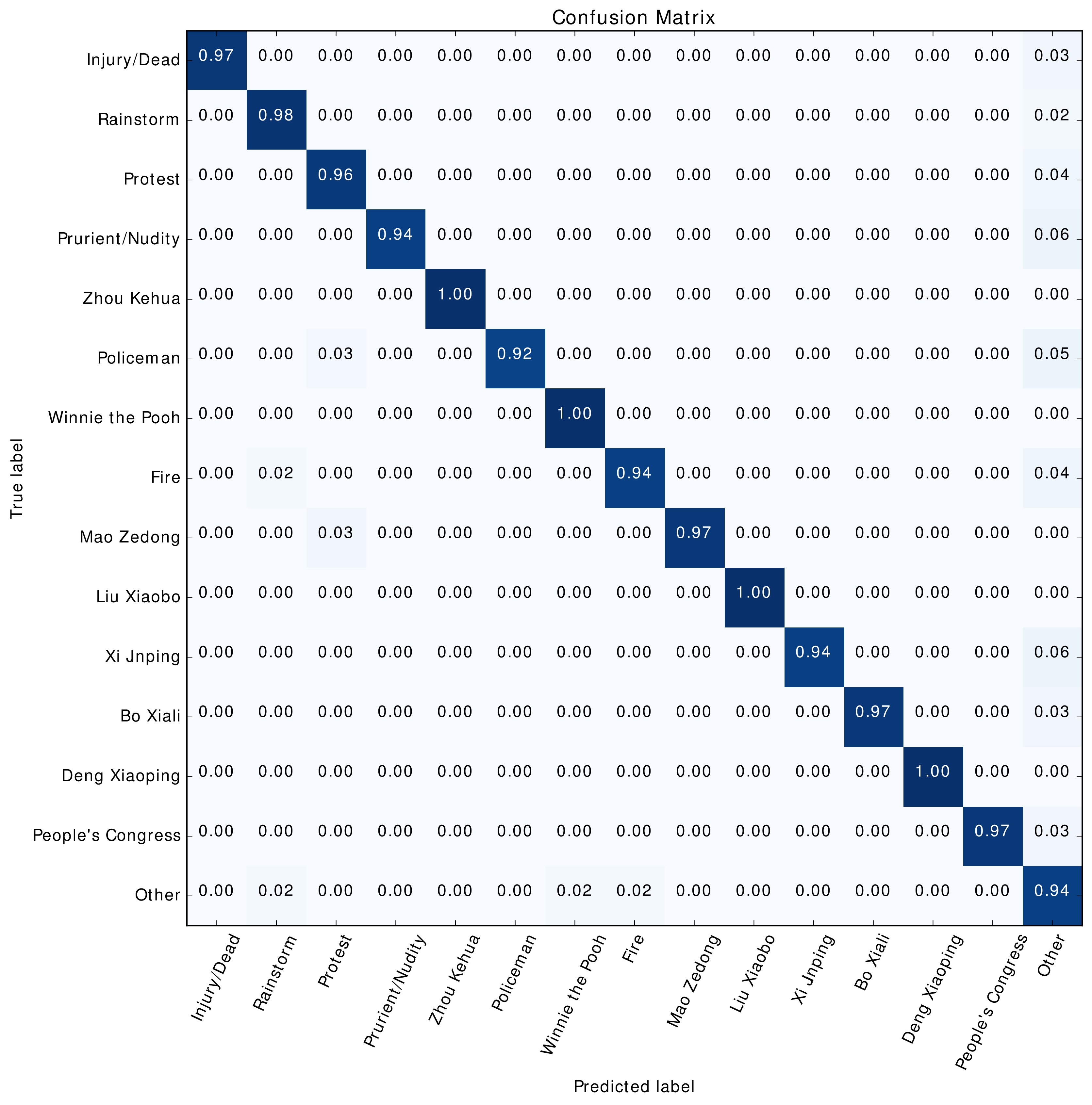}
      \caption{Confusion matrix of image classifier.}
      \label{fig:confusion}
\end{figure}

\noindent\textbf{Classification:} We train a CNN classifier using the VGG-16 network~\cite{simonyan2014very} over the CCTI14's training dataset and then test it with CCTI14's testing dataset. For the training phase, we split the CCTI14's training dataset, stratified by topic, into primary training set (95\% of the data) and development/validation set (5\% of the data). The trained classifier achieves \textbf{97\%} F1-score on the testing dataset. Figure~\ref{fig:confusion} shows the confusion matrix for our classifier. As we can see most of the confusion happens in the ``Other'' class, as expected. 




To reduce the incidence of classifying images that belong to none of our categories as belonging to the most similar category, we used two approaches at the same time: i) \textit{Using an ``Other'' class:} as described in the previous section, ii) \textit{Using a confidence level threshold:} a confidence level threshold of 80\% is used to decide whether to accept the classifier's decision or not, meaning that if the classifier is 80\% or more confident about its decision on an image we accept it, otherwise we categorize it as belonging to the ``Other'' class. We empirically tuned the confidence level threshold on the training data set and achieved the best results with 80\%.


We have evaluated the performance of the classifier using several metrics: precision, recall and F1-score. 
The F1-score takes into account both precision and recall, making it a more reliable metric for evaluating a classifier. The classifier achieves a precision of 97\%, recall of 96\% and  F1-score of 97\% overall.

\subsubsection{Performing CNN Localization}\label{cnn}
To double check our model, we utilized a CNN localization technique introduced by Zhou et al.~\cite{zhou2016learning}. Using the CNN localization technique, we were able to highlight parts of the images that are considered the most important parts by the CNN to decide to classify an image as a specific category. 

We repeatedly used this technique for error analysis and to adjust our model as well as the CCTI14 categories. For example, by highlighting images, we realized that our model was classifying images of Xi Jinping by only looking at his forehead. Although it had a good accuracy in classifying Xi Jinping images, it was suffering from a high false positive rate (23\%), for which the reason was obvious. At that time, this category in CCTI14 had only images from Xi Jinping's face with no background whatsoever.
Therefore, we used the following methodology to mitigate this issue:
\begin{enumerate}
    \item We made an observation such as Xi Jinping is being only classified by his forehead.
    \item We added more diverse images to that category in the CCTI14 dataset to address the problem identified in (1).
    \item If the false positive rate decreased, then we kept the diverse images.
    \item Else, go to (1).
\end{enumerate}
%
%
Following this type of methodology generally for all categories helped us increase the robustness of the trained classifier.

\begin{figure}[t]
\centering
\hspace*{0.0in}%
  \begin{subfigure}{0.24\textwidth}
   \centering
          \fourobjects
  {\includegraphics[width=0.45\textwidth]{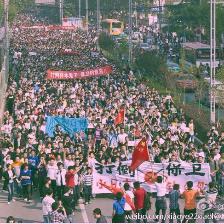}}
  {\includegraphics[width=0.45\textwidth]{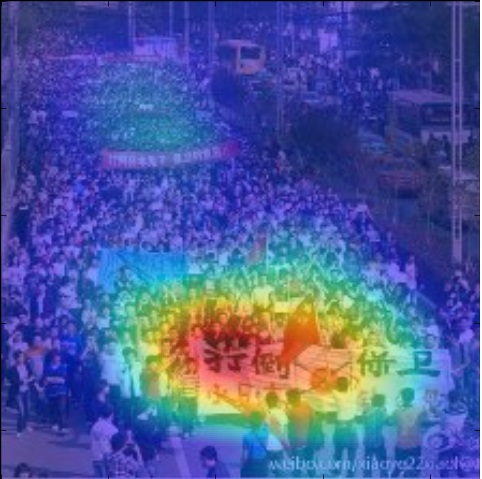}}
  {\includegraphics[width=0.45\textwidth]{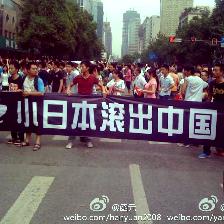}}
  {\includegraphics[width=0.45\textwidth]{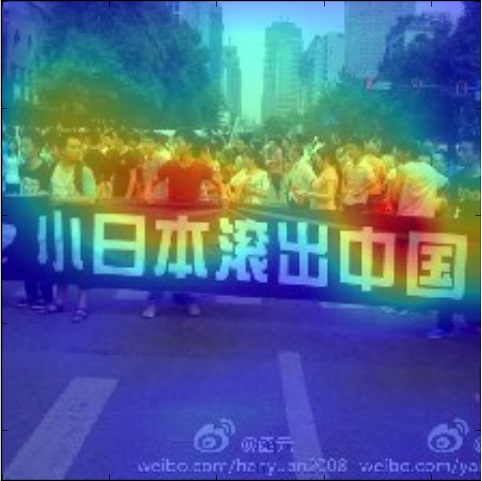}}
    \caption{Protest}
    \label{fig:protest}
  \end{subfigure}
\hspace*{0.05in}%
  \begin{subfigure}{0.24\textwidth}
   \centering
          \fourobjects
  {\includegraphics[width=0.45\textwidth]{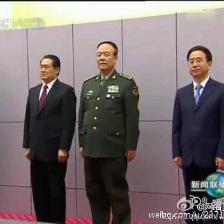}}
  {\includegraphics[width=0.45\textwidth]{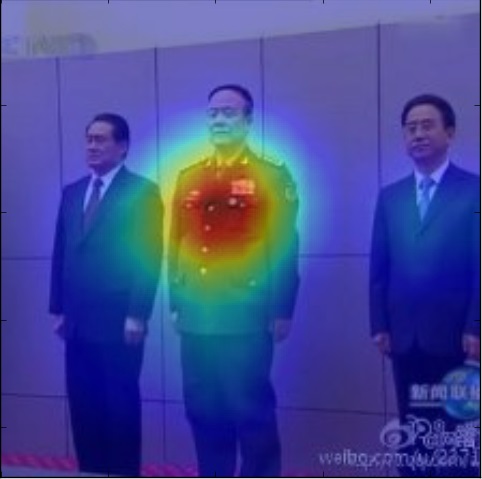}}
  {\includegraphics[width=0.45\textwidth]{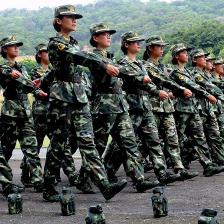}}
  {\includegraphics[width=0.45\textwidth]{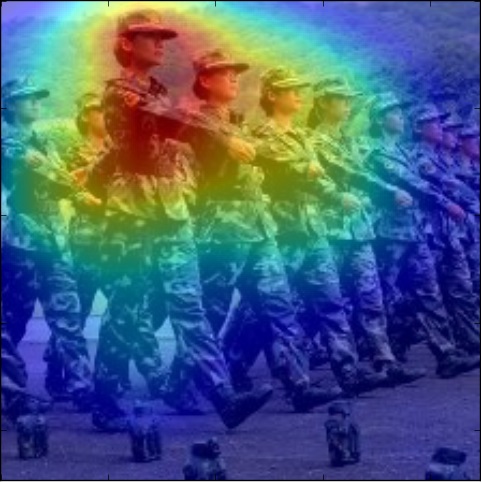}}
    \caption{Policeman}
    \label{fig:policeman}
  \end{subfigure}
\hspace*{0.05in}%
  \begin{subfigure}{0.24\textwidth}
   \centering
          \fourobjects
  {\includegraphics[width=0.45\textwidth]{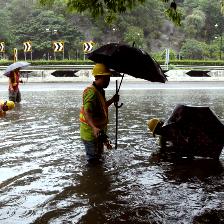}}
  {\includegraphics[width=0.45\textwidth]{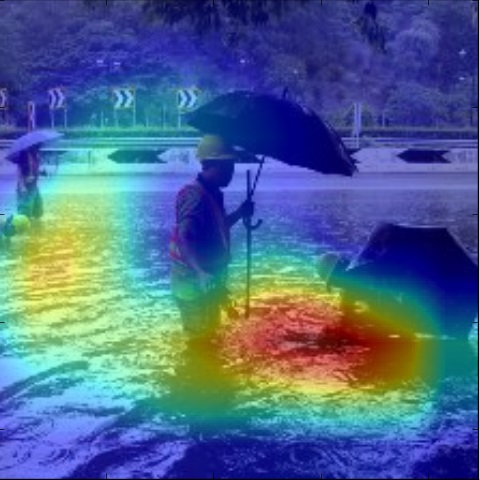}}
  {\includegraphics[width=0.45\textwidth]{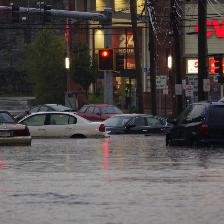}}
  {\includegraphics[width=0.45\textwidth]{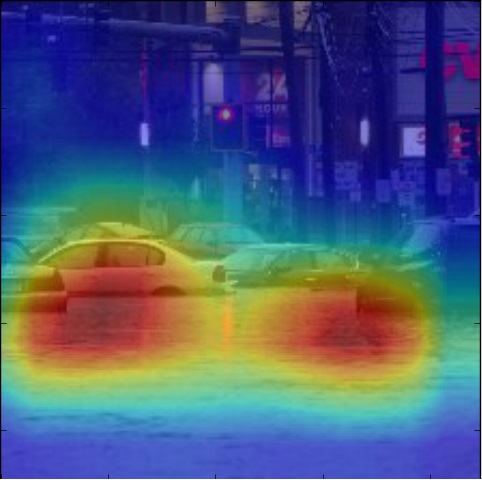}}
    \caption{Rainstorm}
    \label{fig:rainstorm}
  \end{subfigure}
\hspace*{0.05in}%
  \begin{subfigure}{0.24\textwidth}
   \centering
          \fourobjects
  {\includegraphics[width=0.45\textwidth]{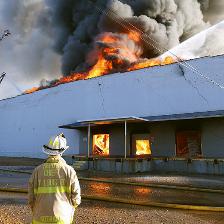}}
  {\includegraphics[width=0.45\textwidth]{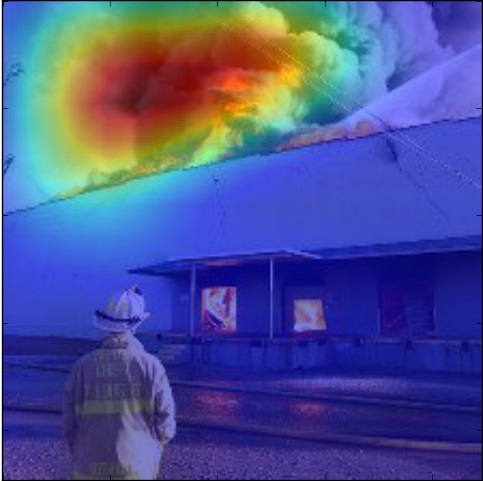}}
  {\includegraphics[width=0.45\textwidth]{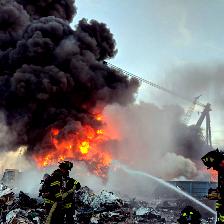}}
  {\includegraphics[width=0.45\textwidth]{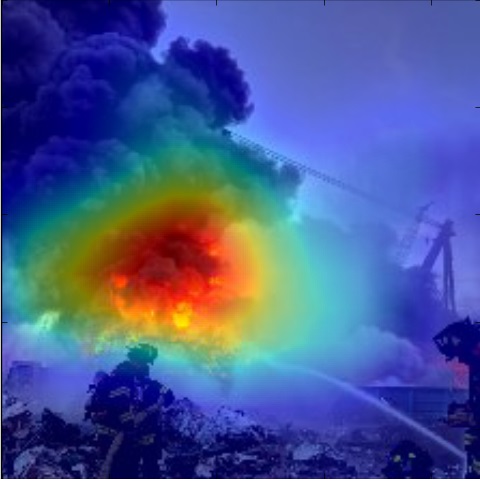}}
    \caption{Fire}
    \label{fig:fire}
  \end{subfigure}
  \caption{Examples of highlighted images.}
  \label{fig:examples}
\end{figure}

Figure~\ref{fig:examples} shows some instances of highlighted images for a few categories. All highlighted parts matched our intuition for each category. 


\begin{figure}[t]
\hspace*{0.0in}%
  \begin{subfigure}{0.24\textwidth}
  \centering
          \fourobjects
  {\includegraphics[width=0.45\textwidth]{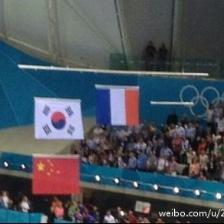}}
  {\includegraphics[width=0.45\textwidth]{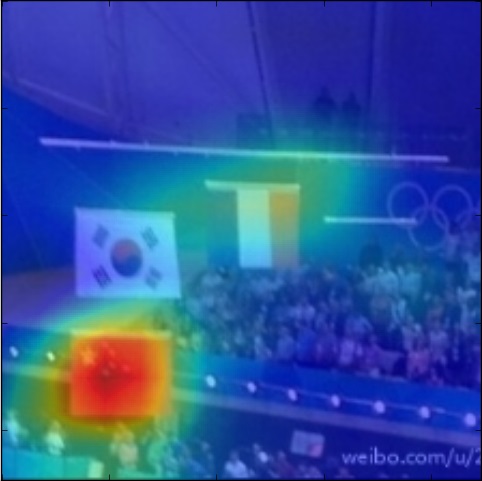}}
  {\includegraphics[width=0.45\textwidth]{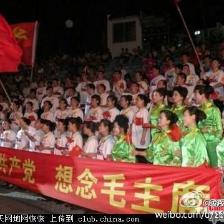}}
  {\includegraphics[width=0.45\textwidth]{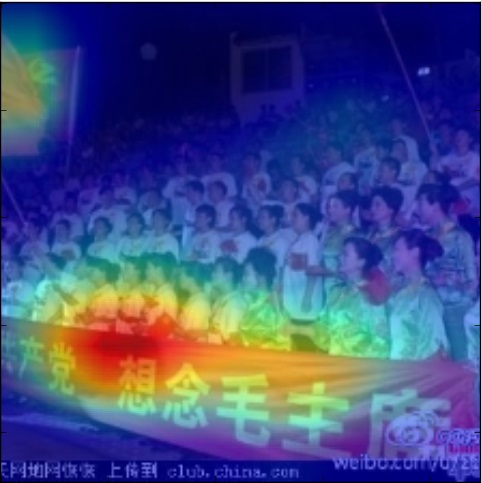}}
    \caption{Protest}
    \label{fig:protest-fp}
  \end{subfigure}
\hspace*{0.05in}%
  \begin{subfigure}{0.24\textwidth}
  \centering
          \fourobjects
  {\includegraphics[width=0.45\textwidth]{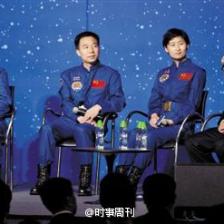}}
  {\includegraphics[width=0.45\textwidth]{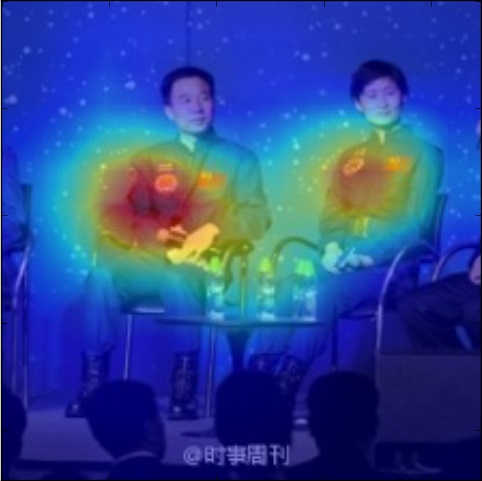}}
  {\includegraphics[width=0.45\textwidth]{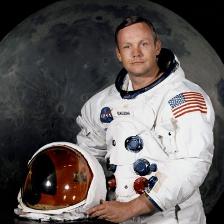}}
  {\includegraphics[width=0.45\textwidth]{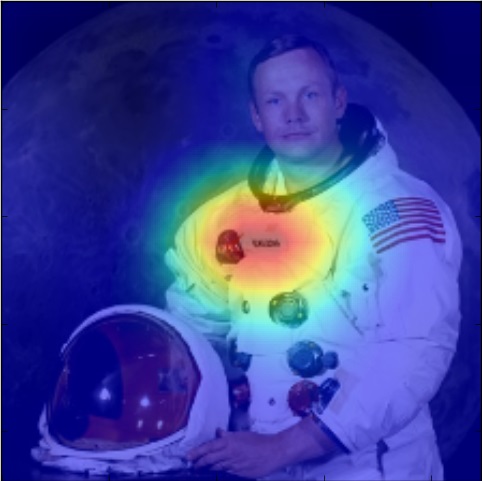}}
    \caption{Policeman}
    \label{fig:policeman-fp}
  \end{subfigure}
\hspace*{0.05in}%
  \begin{subfigure}{0.24\textwidth}
  \centering
          \fourobjects
  {\includegraphics[width=0.45\textwidth]{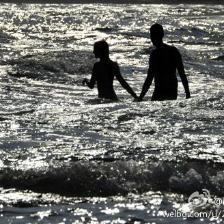}}
  {\includegraphics[width=0.45\textwidth]{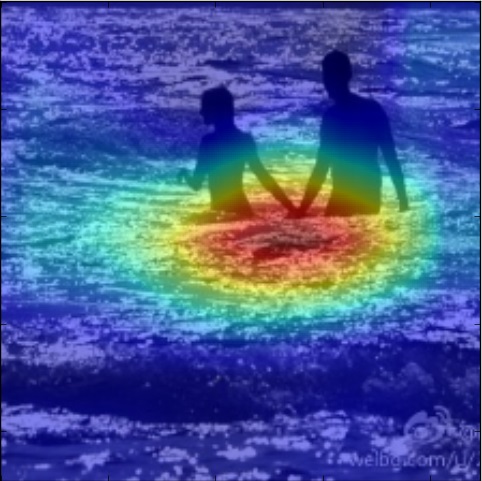}}
  {\includegraphics[width=0.45\textwidth]{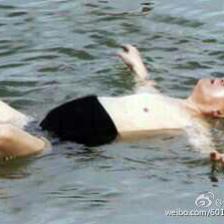}}
  {\includegraphics[width=0.45\textwidth]{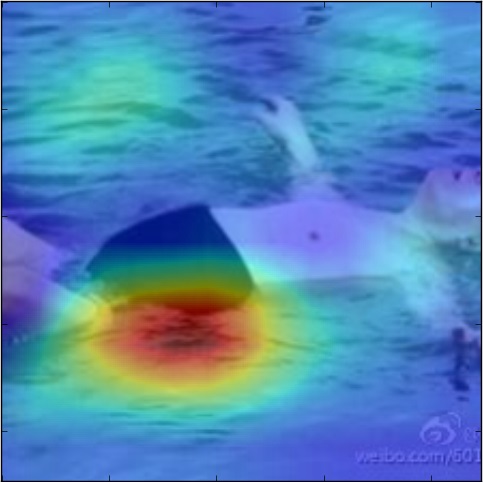}}
    \caption{Rainstorm}
    \label{fig:rainstorm-fp}
  \end{subfigure}
\hspace*{0.05in}%
  \begin{subfigure}{0.24\textwidth}
  \centering
          \fourobjects
  {\includegraphics[width=0.45\textwidth]{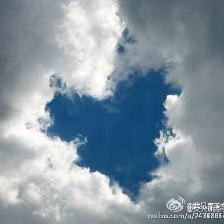}}
  {\includegraphics[width=0.45\textwidth]{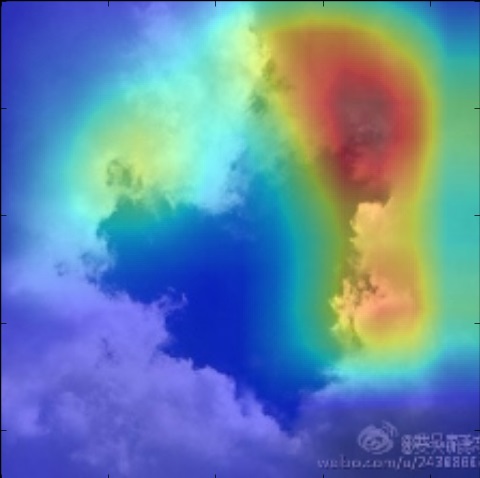}}
  {\includegraphics[width=0.45\textwidth]{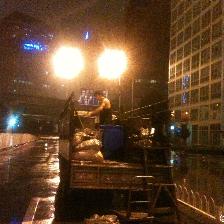}}
  {\includegraphics[width=0.45\textwidth]{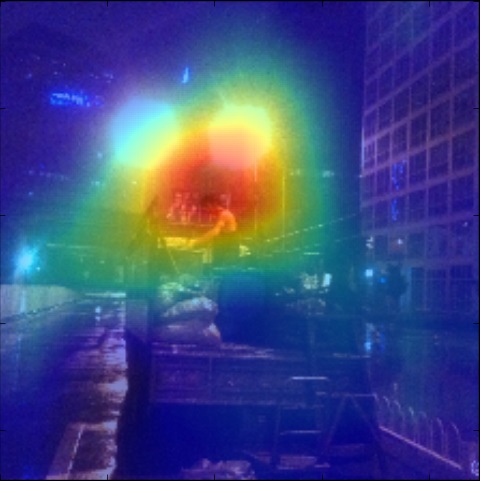}}
    \caption{Fire}
    \label{fig:fire-fp}
  \end{subfigure}
  \caption{Examples of false positives.}
  \label{fig:examples-fp}
\end{figure}

Highlighted examples in Figure~\ref{fig:examples} confirm that our model is trained to look for the right objects in each category. However, some similar objects still can confuse the classifier.
Figure~\ref{fig:examples-fp} shows some examples of the false positives in our model. Images containing something similar to the main features of each category have been incorrectly categorized as that category. 


However, before we do any analysis on the categorized images we manually remove false positives from the 14 categories. Since removing false positives from image categories is fairly easy and it's not very time-consuming, we opt to do so to make our categorized data even more clean.


\subsection{Text Classifier}
To be able to categorize text content of posts into our 14 categories, we built a text classifier. To train our classifier we assembled our own text dataset from \textit{real-world} Weibo posts that we refer to as \textit{CCTT14}. In below we explain how we assembled CCTT14 and then we describe the performance of our text classifier.

\subsubsection{CCTT14 Dataset}

We assembled a text dataset from \textit{real-world} Weibo posts from the same 14 categories as CCTT14 as well as an ``Other'' category, that we refer to as \textit{CCTT14}. Here is the steps we took to assemble this dataset:
\begin{enumerate}
    \item We first trained two human annotators that were native Chinese speakers by providing them the definition of each category as well as examples of each category.
    \item We then partitioned all posts in the WeiboScope dataset using keywords related to each category. We used the keywords extracted by Knockel et al. \cite{knockel2015every} from four Chinese applications as well as the keywords provided by other online resources \cite{words}. The goal of this step was to make the manual annotation process more efficient and less time consuming.
    \item We randomly selected 1000 posts from the output of the previous step.
    \item We asked the two trained annotators to annotate the selected 1000 posts.
    \item We only kept posts that both annotators agreed on their category and if each category had at least 50 posts, we stopped. Otherwise, go to \#3.
\end{enumerate}

The final dataset has 994 labelled posts in which each category has 50-90 posts. Each annotator spent about 12 hours on the whole process, and the inter-reliability of raters was 76\%, which was satisfactory.

\subsubsection{Classifier performance}
\begin{figure}[!ht]
\centering
      \includegraphics[width=0.55\textwidth]{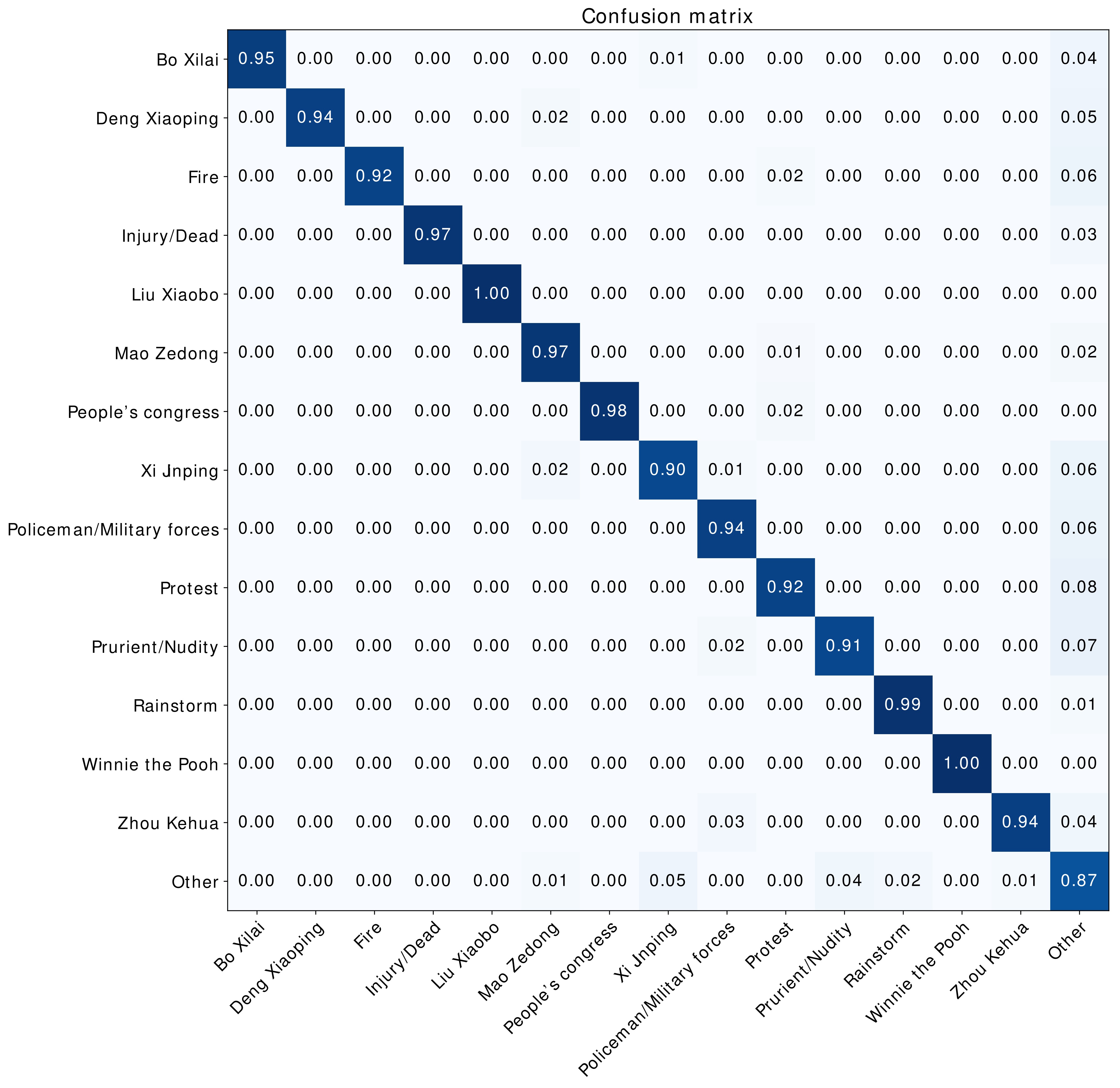}
      \caption{Confusion matrix of the text classifier.}
      \label{fig:confusion-text}
\end{figure}

 We tried different text classifiers (\emph{e.g.}, naive bayes, random forest, neural networks) over CCTT14 and achieved the highest F1-score with multinomial logistic regression. We leveraged unigrams, bigrams, and trigrams as the feature vectors. We also used CoreNLP~\cite{manning2014stanford} tool for word segmentation and tokenization. The classifier achieves a precision of 96\%, recall of 94\%, and F1-score of 95\% overall when we tested our classifier using 10-fold cross validation. Figure~\ref{fig:confusion-text} presents the confusion matrix of our classifier.

\section{Analysis and Results}\label{sec:analysis}
In this section, we present our results on the WeiboScope dataset. We used our classifiers to categorize censored and uncensored posts into our 14 categories and then performed our analysis on the result. 
11,269 censored posts and 7,987 uncensored posts were categorized into the 14 categories.





\subsection{Censorship Rate}
To discover how often a category is censored and what percent of posts in each category is censored, we compared the number of posts found in that category within the censored posts with that of those within the uncensored posts.
Table~\ref{rate-both} shows the number of posts found in each category as well as the percentage of posts in each category that was censored. A post ends up in a category if it has either an image or text in the category.
As one can see in this table, most categories (\emph{e.g.}, protest) are often censored, whereas some categories (\emph{e.g.}, rainstorm) are less frequently censored. This confirms that the consistency of censorship varies by topic/category.  For example, more sensitive categories may experience a higher deletion rate.


\begin{table}
\centering
\ra{1.3}
    \begin{tabular}{lp{1.5cm}p{1.5cm}p{1.5cm}}
    \hline
    Category & \#Censored posts & \#Uncensored posts & Censorship Rate\\\hline
    \textbf{Bo Xilai} & 665 & 336 & 64\%\\ 
    \textbf{Deng Xiaoping} & 281 & 125 & 70\%\\
    \textbf{Fire} & 431 & 530 & 45\%\\ 
    \textbf{Injury/Dead Body} & 1799 & 1029 & 51\%\\ 
    \textbf{Liu Xiaobo} & 184 & 123 &60\%\\ 
    \textbf{Mao Zedong} & 1093 & 486 & 70\%\\ 
    \textbf{People's Congress} & 145 & 113 & 56\%\\
    \textbf{Policeman} & 1311 & 927 & 59\%\\ 
    \textbf{Protest} & 536 & 220 & 71\%\\ 
    \textbf{Prurient/Nudity} & 2664 & 2551 & 51\%\\ 
    \textbf{Rainstorm} & 153 & 207 & 43\%\\ 
    \textbf{Winnie the Pooh} & 160 & 177 & 48\%\\ 
    \textbf{Xi Jinping} & 1745 & 1029 & 63\%\\ 
    \textbf{Zhou Kehua} & 102 & 134 & 43\%\\ 
    \hline
\end{tabular}
    \caption{Percentage of censored posts per category.}
    \label{rate-both}
\end{table}

\subsection{Life Time}
\begin{figure}[!h]
\centering
      \includegraphics[width=0.55\textwidth]{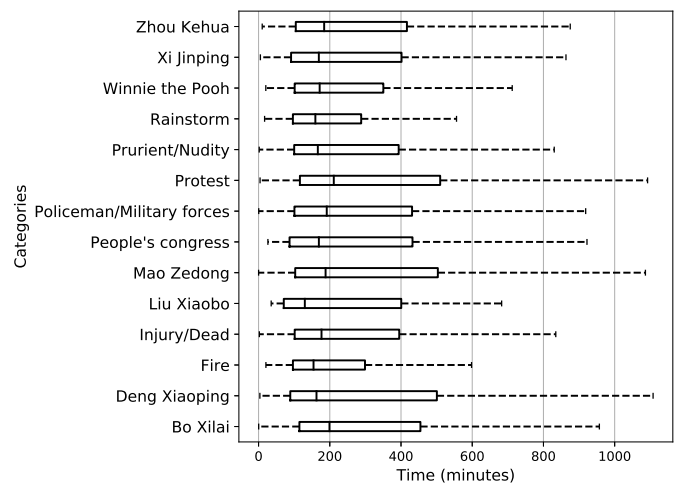}
      \caption{Categories vs.\@ life time}
      \label{fig:lifetime}
\end{figure}

To reveal how quickly posts in a category are censored, we plotted the lifetime distribution of censored posts in that category in minutes. Lifetime is measured as the difference between the time a post is created and the time it is deleted.
Figure~\ref{fig:lifetime} presents the lifetime distribution per category. As one can see, the median lifetime for all categories is less than 180 minutes, meaning that most of the posts are censored in less than three hours. 


\subsection{Survival Analysis}
Survival analysis is used for analyzing data where the outcome variable is the time until an event of interest happens. For example, if the event of interest is death, then the survival time is the time in years, weeks or minutes, etc. until a person dies. In our case, the event of interest is being censored, then the survival time for a post is the time until it is censored. In addition, in survival analysis there are two types of observations: i) those that the event of interest happens during the time of observation (censored posts in our case), ii) those that the event of interest does not happen during the time of observation (uncensored posts in our case). That enables us to take into account both censored and uncensored posts into consideration, despite other researchers that have only considered the censored posts~\cite{zhu2013velocity}.

To analyze how different factors interact to affect censorship, we performed a survival analysis per category over the following measured factors: i) whether the image matches this category, ii) whether the text matches this category, iii) number of reposts, iv) number of comments, and v) text sentiment. To compute the sentiment score we utilized CoreNLP~\cite{manning2014stanford} tool that supports Chinese.


Table~\ref{survival} shows the results of survival analysis per category.
Coefficients in survival analysis relate to hazard (risk of dying or risk of being censored in our case). A positive coefficient for Image, Text, \#Repost, and \#Comment variables means more risk of getting censored and thus shorter lifetime. For example, almost all of the ``Image'' variables have positive coefficient which means having an image that matches that category increases the risk of being censored and therefore shorter lifetime. On the other hand, sentiment is a score between 0-4 (0 being very negative and 4 being very positive). A negative coefficient for sentiment means as we increase the sentiment score (\emph{i.e.} being more positive), it decreases the risk of being censored and therefore longer lifetime.

As shown in Table~\ref{survival}, sentiment always has a negative sign and it is always statistically significant at 5\%. That suggests that \textbf{sentiment is the strongest indicator of censorship across all categories}. Our finding matches with recently leaked logs from Weibo that they were asked by the government to remove all posts about an specific incident, but Weibo advised its censorship department to only deal with the negative content~\cite{blake2019}.

It is also interesting that image category almost always has a positive sign which suggests that having an image that matches that category increases the risk of censorship, but sometimes it is not statistically significant and thus we can not draw firm conclusions about the image category.

\begin{table*}
\centering
\ra{1.3}
    \resizebox{0.95\textwidth}{!}{\begin{tabular}{@{}p{2.6cm}rrrrrrrrrr@{}}\hline
    \multirow{2}{*}{Category} & \multicolumn{2}{c}{Image} & %
        \multicolumn{2}{c}{Text} & \multicolumn{2}{c}{\#Repost} & \multicolumn{2}{c}{\#Comment} & \multicolumn{2}{c}{Sentiment}\\
    \cmidrule(l){2-3}\cmidrule(l){4-5}\cmidrule(l){6-7} \cmidrule(l){8-9}\cmidrule(l){10-11}
     & Coef. & P & Coef. & P & Coef. & P & Coef. & P & Coef. & P\\
    \hline
    \textbf{Bo Xilai} & 0.19 & \textless 0.005 & 0.14 & 0.41 & 0.00 & 0.01 & 0.00 & 0.28 & \textbf{-0.20} & 0.04\\ 
    \textbf{Deng Xiaoping} & 0.62 & 0.01 & 0.04 & 0.87 & 0.00 & 0.41 & 0.00 & 0.52 & \textbf{-0.23} & \textless 0.005\\
    \textbf{Fire} & 0.73 & \textless 0.005 & 0.13 & 0.59 & 0.00 & 0.01 & 0.00 & \textless 0.005 & \textbf{-0.11} & 0.02\\ 
    \textbf{Injury/Dead Body} & 0.63 & 0.02 & -0.02 & 0.94 & 0.00 & 0.12 & 0.00 & 0.54 & \textbf{-0.24} & \textless 0.005\\ 
    \textbf{Liu Xiaobo} & 0.25 & 0.19 & -0.07 & 0.12 & 0.00 & 0.13 & 0.00 & 0.24 & \textbf{-0.27} & 0.04\\ 
    \textbf{Mao Zedong} & 0.31 & 0.09 & 0.05 & 0.85 & 0.00 & 0.40 & 0.00 & 0.04 & \textbf{-0.28} & 0.01\\ 
    \textbf{People's Congress} & 0.16 & 0.07 & -0.21 & 0.18 & 0.00 & 0.03 & 0.00 & 0.34 & \textbf{-0.47} & \textless 0.005\\
    \textbf{Policeman} & 0.19 & 0.24 & 0.09 & 0.62 & 0.00 & \textless 0.005 & 0.00 & 0.36 & \textbf{-0.15} & 0.05\\ 
    \textbf{Protest} & 0.78 & \textless 0.005 & -0.25 & 0.28 & 0.00 & 0.27 & 0.00 & 0.29 & \textbf{-0.06} & 0.05\\ 
    \textbf{Prurient} & 0.74 & \textless 0.005 & 0.09 & 0.68 & 0.00 & \textless 0.005 & 0.00 &  0.19 & \textbf{-0.20} & \textless 0.005\\ 
    \textbf{Rainstorm} & -0.50 & 0.48 & -0.87 & 0.25 & 0.00 & 0.19 & 0.00 & 0.01 & \textbf{-0.31} & 0.02\\ 
    \textbf{Winnie the Pooh} & 0.44 & 0.09 & -0.16 & 0.14 & 0.00 & 0.03 & 0.00 & 0.16 & \textbf{-0.35} & \textless 0.005\\ 
    \textbf{Xi Jinping} & 0.49 & \textless 0.005 & -0.51 & \textless 0.005 & 0.00 & 0.74 & 0.00 & 0.07 & \textbf{-0.09} & 0.01\\ 
    \textbf{Zhou Kehua} & 0.22 & \textless 0.005 & -0.08 & 0.11 & 0.00 & 0.04 & 0.00 & 0.23 & \textbf{-0.17} & \textless 0.005 \\\hline
\end{tabular}}
    \caption{Survival regression per category.}
    \label{survival}
\end{table*}

\section{Related Work}




The Weibo platform is popular and previous researchers have attempted to study its censorship mechanism. King \emph{et al.}~\cite{king2013censorship} collected a dataset of censored posts, by checking for the deleted posts \textit{every 24 hours}, over \textit{six months} in 2011. Using that dataset, they identified the collective action potential of posts as a major indicator of censorship. Bamman \emph{et al.}~\cite{bamman2012censorship} used a dataset collected over \textit{three months} in 2011, and performed a statistical analysis of deleted posts and showed that posts with some sensitive words are more likely to be deleted. Zhu \emph{et al.}~\cite{zhu2013velocity} collected a dataset of censored posts by tracking \textit{3,567 users} over \textit{three months} in 2012. They investigated how quickly, on a scale of minutes, posts in Weibo are removed. They also performed a logistic regression over \textit{censored data only} to analyze the interaction of different factors, \textit{by ignoring sentiment and topics}, and showed that whether a post contains an image has the highest impact on censorship. 


Ng \emph{et al.}~\cite{ng2018detecting} built a Naive Bayes classifier over \textit{344} censored and uncensored posts related to Bo Xilai scandal to predict censorship. They indicated that posts with subjective content, \emph{e.g.} expressions of mood and feeling, are likely to be censored. Ng \emph{et al.}~\cite{ng2018linguistic} collected \textit{2,171} censored and uncensored posts from 7 categories and built a text classifier based on linguistic features (\emph{e.g.}, sentiment) to predict censorship. They indicated that the strongest linguistic feature in censored posts is readability.

However, no study to date has considered the multimodal nature of social
media, and past studies have relied on relatively narrow datatsets
(\emph{e.g.}, spanning months rather than years or only following a
small set of users).




To the best of our knowledge there is no prior work that has studied image censorship in Weibo, but there are some previous work that have studied \textit{WeChat}, the most popular chat application in China.
Crete-Nishihata \emph{et al.}~\cite{lx} discovered WeChat image censorship for the first time in individual chats, in addition to group chats after the death of Liu Xiaobo. Recently Knockel \emph{et al.}~\cite{knockel2018analysis} investigated image censorship mechanisms in WeChat. They indicated that WeChat utilizes two image filtering algorithms: i) an Optical
Character Recognition (OCR)-based algorithm to censor sensitive text, ii) a visual-based algorithm to censor images visually similar to images in an image blacklist.

\section{Conclusion}

In this paper, we analyzed a dataset of censored and uncensored posts from Weibo using deep learning, NLP techniques, and manual effort. We first introduced the CCTI14 and CCTT14 datasets with 14 categories designed particularly for studying image and text censorship in China. Then we trained classifiers on CCTI14 and CCTT14 and used the classifiers to classify the target dataset so that we can analyze censorship mechanisms in Weibo. 

Using our classifiers, we found that \textit{sentiment} is the only indicator of censorship that is consistent across the variety of topics we identified. Our finding matches with recently leaked logs from Weibo. We also found that some categories (\emph{e.g.}, protest) are often censored, while some categories (\emph{e.g.}, rainstorm) are less frequently censored.  Our analysis suggests that all the posts from our 14 categories are deleted in less than three hours on average, which confirms that censors can delete sensitive content very quickly.  Taken as a whole and within the body of other related research, our results call into question the idea that censorship are binary decisions devoid of timing or context.  The ``there are a set of sensitive topics and any content within that set are censored'' view of censorship needs to be reevaluated.

We believe that CCTI14 and CCTT14 could be an essential first step towards analyzing image and text censorship in different platforms in China. We make CCTIT14 and CCTT14 along with our trained models publicly available so that it can help efforts to analyze similar tasks in other Chinese platforms. 


\section*{Acknowledgments}

This research has been supported by the U.S.\ National
Science Foundation (Grant No. \#1518918).

\Urlmuskip=0mu plus 1mu\relax
\bibliographystyle{acm}
\bibliography{arxiv2019}

\end{document}